\documentclass[twocolumn]{aastex63}
\usepackage{xcolor}
\usepackage{amsmath}
\usepackage{amssymb}
\usepackage{graphicx}

\definecolor{darkgreen}{rgb}{0,0.6,0.2}

\definecolor{lightblue}{rgb}{0.19, 0.55, 0.91}

\definecolor{amethyst}{rgb}{0.6,0.4,0.8}

\newcommand{\fGW}{\ensuremath f_{\rm GW}}
\newcommand{\dotfGW}{\ensuremath {\dot f}_{\rm GW}}

\newcommand{\CCA}{\affiliation{Center for Computational Astrophysics, Flatiron Institute, 162 5th Ave, New York, NY 10010, USA}}
\newcommand{\CIT}{\affiliation{Department of Physics, California Institute of Technology, Pasadena, California 91125, USA}}
\newcommand{\CITLab}{\affiliation{LIGO Laboratory, California Institute of Technology, Pasadena, CA 91125, USA}}

\received{\today}
\revised{}
\accepted{}
\submitjournal{ApJ}

\newcommand{\referee}[1]{#1}

\graphicspath{{./}{figures/}}

\begin{document}

\title{Prospects of gravitational-waves detections from common-envelope evolution with LISA}
\shorttitle{GWs from common envelope with LISA}

\correspondingauthor{M.~Renzo}
\email{mrenzo@flatironinstitute.org}

\author[0000-0002-6718-9472]{M.~Renzo}
\affiliation{Department of Physics, Columbia University, New York, NY 10027, USA}
\CCA
\author[0000-0001-9892-177X]{T.~Callister}
\CCA
\author[0000-0002-5833-413X]{K.~Chatziioannou}
\CIT
\CITLab
\CCA
\author[0000-0001-5484-4987]{L.~A.~C.~van~Son}
\affiliation{Center for Astrophysics | Harvard \& Smithsonian, 60 Garden Street, Cambridge, MA 02138, USA}
\author[0000-0002-4307-1322]{C.~M.~F.~Mingarelli}
\affiliation{Department of Physics, University of Connecticut, 196 Auditorium Road, U-3046, Storrs, CT 06269-3046, USA} \CCA
\author[0000-0002-8171-8596]{M.~Cantiello} \CCA
\affiliation{Department of Astrophysical Sciences, Princeton University, Princeton, NJ 08544, USA}
\author[0000-0002-5956-851X]{K.~E.~S.~Ford}
\affiliation{Dept. of Science, Borough of Manhattan Community College, City University of New York, New York, NY 10007 USA}
\affiliation{Dept. of Astrophysics, American Museum of Natural History, New York, NY 10024 USA}
\affiliation{Physics Program, CUNY Graduate Center, City University of New York, New York, NY 10016 USA}
\CCA
\author[0000-0002-9726-0508]{B.~McKernan}
\affiliation{Dept. of Science, Borough of Manhattan Community College, City University of New York, New York, NY 10007 USA}
\affiliation{Dept. of Astrophysics, American Museum of Natural History, New York, NY 10024 USA}
\affiliation{Physics Program, CUNY Graduate Center, City University of New York, New York, NY 10016 USA}
\CCA
\author[0000-0001-7288-2231]{G.~Ashton}
\affiliation{Royal Holloway University of London, Egham, Surrey, TW20 0EX, U.K.}
\affiliation{School of Physics and Astronomy, Monash University, VIC 3800, Australia}

\begin{abstract}
  Understanding common envelope (CE) evolution is an outstanding
  problem in binary evolution. Although the CE phase is not driven by
  gravitational-wave (GW) emission, the in-spiraling binary emits GWs
  that passively trace the CE dynamics. Detecting this GW signal would
  provide direct insight on the gas-driven physics. Even a
  non-detection might offer invaluable constraints. We investigate the
  prospects of detection of a Galactic CE by LISA. While the dynamical
  phase of the CE is likely sufficiently loud for detection, it
  is short and thus rare. We focus instead on the
  self-regulated phase that proceeds on a thermal timescale. Based on
  population synthesis calculations and the (unknown) signal
  duration in the LISA band, we expect $\sim 0.1-100$ sources in the
  Galaxy during the mission duration. We map the GW observable
  parameter space of frequency $f_\mathrm{GW}$ and its derivative
  $\dot f_\mathrm{GW}$ remaining agnostic on the specifics of the
  inspiral, and find that signals with $\mathrm{SNR}>10$ are possible
  if the CE stalls at separations such that
  $f_\mathrm{GW}\gtrsim2\times10^{-3}\,\mathrm{Hz}$. We investigate
  the possibility of mis-identifying the signal with
  other known sources. If the second derivative
  $\ddot f_\mathrm{GW}$ can also be measured, the signal can be
  distinguished from other sources using a GW braking-index.
  Alternatively, coupling LISA with electromagnetic observations of
  peculiar red giant stars and/or infrared and optical transients,
  might allow for the disentangling of a Galactic CE from other
  Galactic and extra-galactic GW sources.
\end{abstract}
\vspace*{-20pt}
\keywords{Common-envelope evolution --- Common-envelope binary stars
  --- Gravitational-wave astronomy}

\section{Introduction}
Common envelope (CE) evolution is one of the most challenging
processes to study in the evolution of binary stars.
CE evolution was originally proposed by
\cite{paczynski:76} to account for the short orbital separations
observed in cataclysmic variables.
Despite its
importance for many categories of stellar objects,
the details of this process remain elusive.

A CE is typically assumed to occur when the evolution of a binary
system leads to dynamically unstable mass transfer, causing the
accretor to be engulfed in the donor's envelope. This results in a CE
binary, in which the core of the donor star and the accretor star
coexist within a shared envelope. Drag forces then drive the inspiral
of the binary, injecting orbital energy \citep[and angular momentum,
][]{nelemans:00} into the CE. Depending on the complex coupling
between the orbit and the envelope structure, the CE can ultimately be
ejected, or, alternatively, the two stars can merge.  We expand on the
current understanding of CE evolution in Sec.~\ref{sec:ce_theory}.

Evolution through a CE phase is a key step in the formation of almost
any compact binary system from an isolated binary, including cataclysmic variables
\citep[e.g.,][]{paczynski:76}, subdwarf B-stars \citep[e.g.,][]{han:03,
  igoshev:20} and stripped stars \citep{gotberg:20}, double
white dwarfs \citep[WD, e.g.,][]{nandez:15}, X-ray binaries
\citep[e.g.,][]{podsiadlowski:02, chen:20}, binary neutron-stars
\citep[e.g.,][]{tauris:17, kruckow:18, vigna-gomez:20}, black-hole
neutron-star binaries \citep[e.g.,][]{kruckow:18, chattopadhyay:20, broekgaarden:21},
black-hole WD binaries \citep[e.g.,][]{sberna:20}, and binary
black-holes \citep[e.g.,][]{belczynski:16nat, kruckow:18}. CE evolution
also produces the tightest binaries which
can in turn make the fastest runaway stars from isolated systems if
disrupted by the explosion of one star
\citep[e.g.,][]{justham:09, renzo:19walk, evans:20, neunteufel:20}. Finally, CE preceeds the
majority of stellar mergers\footnote{We neglect here the possibility
  of (close to) head-on collisions between stars in dense stellar environments
  which can also lead to mergers \citep[e.g.][]{glebbek:09, dicarlo:20, renzo:20merger}.}, which
are common across the entire mass range of stars
\citep[e.g.,][]{demink:13,zapartas:17,temmink:20} and may possibly
lead to the formation of Thorne-Zytkow objects
\citep[e.g.,][]{thorne:77}.

The final phases of a CE -- the envelope ejection or the merger event
-- can produce visible electromagnetic (EM) transients called luminous
red novae (LRN). The detection of the pre-transient period decay of
V1309Sco \citep{tylenda:11} unambiguously connected this type of
transient to CE. LRNe exhibit a variety of morphologies
\citep[e.g.,][]{pastorello:20}, possibly related to the variety of CE
progenitors and outcomes possible.

Numerical simulations of CE evolution remain challenging (e.g.,
\citealt{ricker:08, ohlmann:16, macleod:18,chamandy:20,
  cruz-osorio:20, sand:20}, see also \citealt{ivanova:13, ivanova:20}
for reviews). The main difficulties relate to the large range of
temporal and spatial scales that need to be resolved in CE
evolution. Moreover, there are several possible energy sources other
than the orbit that need to be considered during CE evolution, such as
recombination energy \citep[e.g.,][]{ivanova:13b, ivanova:18, sand:20},
accretion energy
\citep[e.g.,][]{voss:03,macleod:14}, and energy released by nuclear burning
\citep[e.g.,][]{ivanova:02}. Both where the energy is released, and how it
is transported through the CE also have a crucial role in determining
the outcome \citep[e.g.,][]{sand:20, wilson:20}. Therefore, CE
evolution remains one of the major uncertainties in the modeling of
binary evolution and stellar populations.

The inspiral within a CE is not driven by gravitational-wave (GW)
emission. Nevertheless, the binary inside of a CE, consisting of the
core of the donor and the companion star, has a time-varying mass
quadrupole moment and hence radiates GWs.
If detectable, GWs from the binary inside of a CE will passively trace
the gas-drag driven inspiral, and could provide invaluable insight
into the CE process.  Conversely, because of the low density of the
envelope of giant stars (which is possibly further lowered by the energy
injection during the CE), any GW signal from the envelope itself is
unlikely to be detectable.

The GWs radiating from a Galactic CE binary would be in the mHz
band, and thus could be detectable by space-based GW-detectors, such
as the Laser Interferometer Space Antenna (LISA,~\citealt{LISA:2019}),
and TianQin \citep{TianQin}. This can be expected because of the
existence of double-WD verification binaries for LISA \citep[e.g.,][]{stroeer:06}, which likely
went through at least one CE phase \citep[e.g.,][]{korol:17}. LISA is
currently scheduled for launch in 2034 and will commence operations in
the mHz GW frequency range shortly thereafter. Its projected mission
lifetime is $5-10$ years, during which time it is expected to detect a
wide variety of GW signals in the mHz band, from supermassive black
hole binary mergers to Galactic compact binaries with a fraction of a
solar mass.

Here, we examine the prospects for observing the inspiral of two
Galactic stellar cores within a CE with LISA. Such signals are not
accessible to ground-based detectors (e.g., LIGO -- \citealt{LIGO},
Virgo -- \citealt{Virgo}, and KAGRA -- \citealt{somiya:12}) as
they are restricted to frequencies above a few to tens of Hz. At these
frequencies, CE binaries are expected to either have merged or have
ejected the envelope.
In fact, \cite{ginat:20} showed that LISA could detect the portion of
the CE inspiral proceeding on a dynamical timescale ($\sim$
days).

However, not all CE events are expected to rapidly complete
their evolution over a dynamical timescale~\citep{ivanova:13}. A
significant fraction of the envelope may be ejected on thermal
timescales much longer than the dynamical timescale
\citep[e.g.,][]{meyer:79,glanz:18,michaely:19}.  Currently, the
duration and occurrence rate of this so-called `self-regulated
inspiral' is very uncertain \citep[e.g.,][]{meyer:79,fragos:19,sand:20,
  igoshev:20}, but if it is of the order of years or longer, this
phase is likely to constitute the most promising CE target for LISA.

Constraints placed by low-frequency GW on CE evolution will be from
direct detection/non-detection. These constraints will be
complementary to those obtained by ground-based detectors which
observe the aftermath of successful CE events in massive binaries
where the envelope has been ejected and the binary merged under the
influence of GWs.  The properties and rate of such detections can be
used to place constraints on the CE properties, for example the CE
efficiency~\citep[e.g.,][]{wong:20, zevin:20}. Space-based detectors,
on the other hand, offer the possibility of studying binaries while
they undergo CE, possibly including systems that might fail to eject
the envelope. Provided the signal can be distinguished from other
potential mHz GW sources, space-based detectors will give us direct
insight into the gas-driven dynamics.

In particular, the GW frequency
of the signal ($f_\mathrm{GW}$), can constrain the orbital separation at
which the self-regulated inspiral stalls, and the frequency change
($\dot{f}_\mathrm{GW}$) and event duration can constrain the rapidity of
the inspiral and thus the internal gas density of the CE (and
potentially its radial distribution).  Even a non-detection would
offer invaluable constraints on the occurrence rate, the orbital
separation during the self-regulated inspiral, and the duration of
this phase, which are not accessible by any other means.

\subsection{Phases of common envelope evolution}
\label{sec:ce_theory}

\begin{figure}[tbp]
  \centering
  \includegraphics[width=0.5\textwidth]{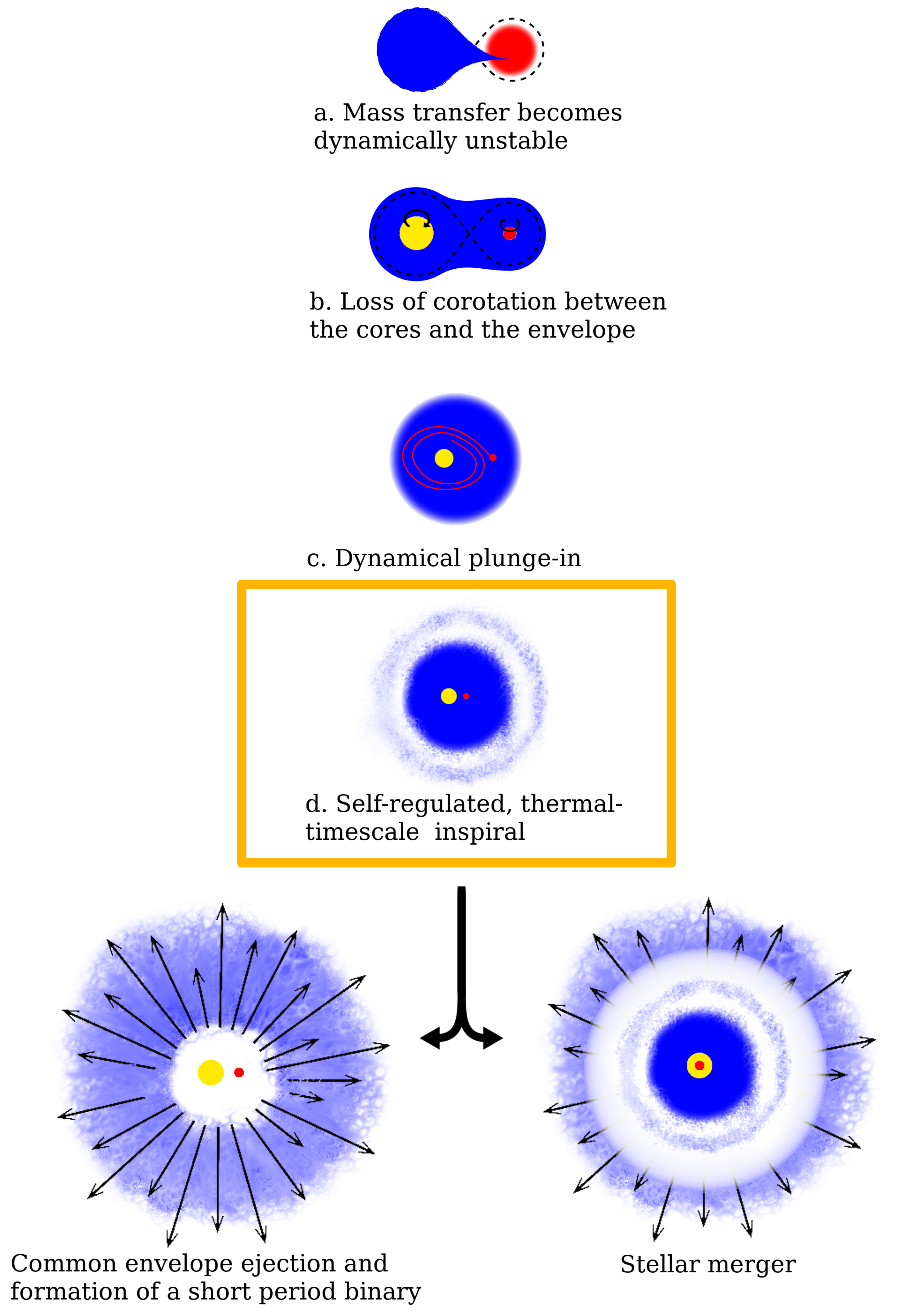}
  \caption{Schematic representaion of all possible phases during a common envelope
    event. We focus on the most promising source of GW in the LISA
    band, i.e. the thermal-timescale self-regulated phase at step
    d. However, dynamical phases such as the plunge-in at step c or
    the final ejection or merger might also be detectable \citep[e.g.,][]{ginat:20}.}
  \label{fig:cartoonCE}
\end{figure}

Because of the aforementioned complexity of CE evolution and the
diversity of binary systems that can evolve through this phase (both
in terms of masses and evolutionary stages at the onset of the CE), a
complete and exhaustive picture of this phenomenon does not yet
exist. Nevertheless, several phases which might subsequently occur
during the CE have been identified \citep[e.g.,][]{podsiadlowski:01,
  ivanova:13, fragos:19, ivanova:20}, and are outlined in
Figure~\ref{fig:cartoonCE}. Below we emphasize which are potentially interesting
targets for GW observations. It is possible that not every binary
evolving through a CE will experience all of these phases
\citep{ivanova:13, lawsmith:20}.

\paragraph{a.~Onset of the common envelope} Mass transfer through a
Roche lobe overflow (RLOF) can become dynamically unstable and result
in the initiation of a CE for several reasons:
\begin{itemize}
\item[(i)]an increasing rate of overflow -- either because of
  the Roche lobe shrinks or because it expands slower than the
  donor's star radius;
\item[(ii)] because the mass transfer timescale is much shorter
  than the accretor thermal timescale, the accretor is pushed out of
  thermal equilibrium. This leads to its radial expansion until the
  accretor too fills its Roche lobe \citep[e.g.,][]{nariai:76};
\item[(iii)] because of the Darwin instability that redistributes angular
  momentum between the stellar spins and the orbit
  \citep{darwin:1879}, causing the orbit to shrink;
\item[(iv)] because of mass loss from the L$_2$ Lagrangian point,
  causing large orbital energy loss. This accelerates the
  evolution to proceed on a dynamical timescale
  \citep[e.g.,][]{pejcha:17}.
\end{itemize}
These processes can cause the accretor to enter the envelope of the
donor, starting the CE on a timescale that is either thermal or
dynamical. Possible pre-CE mass ejection can also occur, depending on
what drives the CE initiation
\citep[e.g.,][]{pejcha:17, macleod:17,macleod:20}. When RLOF becomes dynamically
unstable, the separation, $a$, between the core of the donor and the
accretor is still wide and of the order of the Roche radius of the
dono,r $a\gtrsim R_\mathrm{RL,1}$. Except for binaries already involving
two compact objects \citep[e.g., AM CVn systems,][]{liu:21}, this is likely to correspond to a
GW signal of frequency $f_\mathrm{GW}$ too low for the LISA band.

\paragraph{b.~Loss of co-rotation} Soon after the initiation of the
CE, the envelope will rapidly stop co-rotating with the binary
(even under the optimistic assumption that the envelope was initially co-rotating with
the donor's core). The timescale for the loss of co-rotation depends
on how the CE was initiated at step (\emph{a}), but it varies between the
tidal timescale up to a thermal timescale
\citep[e.g.,][]{podsiadlowski:01, fragos:19}.
The loss of co-rotation still happens at
separations wider than could be detectable in the LISA band, but it
allows for the onset of strong gas-drag which drives the
subsequent CE evolution.

\paragraph{c.~Dynamical plunge-in} After the loss of co-rotation, with
the envelope density still only slightly perturbed \citep{fragos:19},
a phase of dynamical plunge-in happens. This rapid inspiral can take
(a fraction of) an orbital period and possibly decrease the orbital
separation $a$ to zero (i.e., a prompt merger). Alternatively it
can lead to a prompt ejection of the CE
\citep[e.g.,][]{lawsmith:20}. However,
this is not always the case. The orbital energy
liberated in this very fast plunge-in could expand the CE, leading to a
decrease in gas-drag and thus stalling the inspiral, setting up
the conditions for phase (\emph{d}) (see below). At which separation
$a_\mathrm{post-plunge}$ the plunge-in stalls is likely not to be
universal, and is not yet fully understood. \cite{podsiadlowski:01}
suggested that $a_\mathrm{post-plunge}$ corresponds to the location
where the orbital energy equates the envelope binding energy, however
\cite{demarco:11} caution about the subtleties of how the binding
energy is defined. Depending on the pre- and post-plunge-in
separations, this dynamical phase could be detectable in GW for Galactic
binaries \citep[e.g.,][]{ginat:20}. However, because of its short
duration ($\sim$ hours -- days), it is not the most promising source (see
Sec.~\ref{sec:rate}). Should one happen within the Galaxy during the
LISA mission, it would produce a frequency change
$\dot{f}_\mathrm{GW}$ much larger than expected from GW emission, and
possibly be associated to a detectable EM transient. The
plunge-in phase can also produce an eccentric binary within the CE
\citep{ivanova:13, sand:20}, with potential implications for the GW
signal they might produce.

\paragraph{d.~Self-regulated inspiral} Assuming the plunge-in did not
produce a prompt merger, what follows is a relatively slow phase of
self-regulated inspiral, that can last several thermal time-scales
(10-$10^5$\,years, e.g., \citealt{meyer:79, clayton:17, fragos:19, chamandy:20,
  igoshev:20}). During the self-regulated inspiral, the
quasi-Keplerian motion of the binary slowed by gas-drag
progressively injects orbital energy and angular momentum in the
CE. Other energy sources can intervene
and radiative transport becomes relevant during this CE phase
 (e.g., hydrogen recombination can occur if it happens in a sufficiently
optically thick layer, \citealt{fragos:19}). Depending on the separation
$a_\mathrm{post-plunge}$ (and the masses of the binary), this
phase might be the most promising for detecting a GW signal, since it
could allow for a slowly\footnote{We still expect
  $\dot{f}_\mathrm{GW}$ to exceed the value predicted by
  general relativity for GW-driven inspiral in vacuum.} varying $f_\mathrm{GW}$ that can be
integrated over several years to build up signal-to-noise ratio
(SNR). However, we emphasize that this
self-regulated inspiral might not happen for all CE events (see for
example the short timescale for subdwarf B-type progenitors inferred
by \citealt{igoshev:20}, or the numerical results for neutron star
progenitors by \citealt{lawsmith:20}).

\paragraph{e.~CE ejection or merger} Finally, the CE will end with one
last dynamical phase during which either the envelope is ejected
(possibly because of the intervention of energy sources other than the
orbit) or one of the stars of the binary inside the CE
fills the equipotential surface distinguishing it from the other
(improperly, its Roche lobe) leading to the final merger
\citep[e.g,][]{podsiadlowski:01, ivanova:02}. In the cases where the
CE is successfully ejected, the surviving binary is left with the final
$a_\mathrm{post-CE}$ orbital separation smaller than the initial
radius of the donor. What governs the binary post-CE
orbital evolution depends on the specific details of the system (magnetic
braking, tidal evolution, wind mass loss, or possibly GW emission).  The short duration of phase
(\emph{c}), phase makes it an unpromising
target for GW searches. However, it would result in a large
$\dot{f}_\mathrm{GW}$ for a brief moment, and \referee{likely} be
accompanied by a detectable EM transient.\\

In the rest of this study, we will focus on potential
signals from a long lived, self-regulated inspiral (step \emph{d})
unless otherwise stated. In Sec.~\ref{sec:crossing_band} we
show that CE binaries of various mass are expected to cross or end within the LISA GW
frequency band. We then consider the issue of whether a typical binary
crossing the LISA band can be detected in Sec.~\ref{sec:detection}, and use population
synthesis simulations to estimate the number of sources expected at
any given time in the Galaxy in
Sec.~\ref{sec:rate}. Sec.~\ref{sec:bias} deals with the possible
``stealth bias'', that is the possibility of misinterpreting the
detection of GWs from a CE as another kind of GW source, and we
consider how EM observation can help resolving this bias in Sec.~\ref{sec:EM_counterparts}. We further
discuss our results in Sec.~\ref{sec:discussion} before concluding in Sec.~\ref{sec:conclusions}.

\section{Common-envelope evolution across the LISA band}
\label{sec:crossing_band}
\begin{figure}[tbp]
  \centering
  \includegraphics[width=0.45\textwidth]{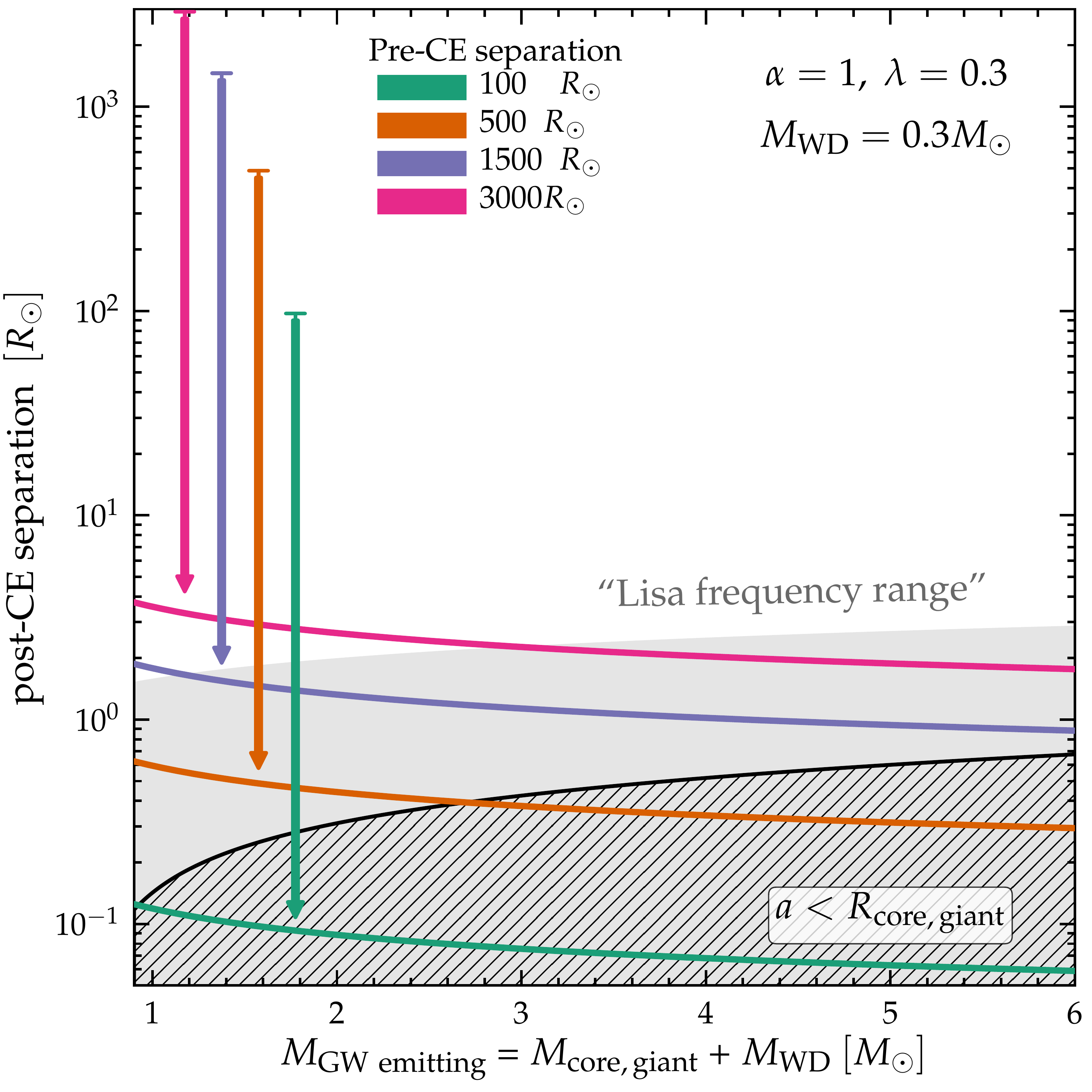}
  \caption{Estimates of the post-CE separations using the
    $\alpha\lambda$ energy formalism. Given a pre-CE separation
    (different colors in plot) and the GWs emitting mass (x-axis), the
    post-CE separation can be determined. We show a range of
    GW-emitting masses, consisting of the core mass of the donor star
    plus the mass of the companion WD (0.3$\,M_\odot$ in this
    plot). We assume $\alpha = 1$ and $\lambda = 0.3$ for illustrative
    purposes. The gray band shows where the corresponding GW signal
    approximately be in the LISA frequency range
    ($10^{-4}\,\mathrm{Hz}\lesssim f_\mathrm{GW}\lesssim 0.1\,\mathrm{Hz}$). Systems
    with post-CE separations within or below this band therefore
    potentially experience a phase during which their GW emission
    could be detectable by LISA. The hatched area corresponds to
    post-CE separation smaller than the core radius, that is presumably
    merging systems.}
  \label{fig:CE_sep}
\end{figure}

For a CE event to cross the LISA frequency band and produce a
detectable GW signal, the mass of the binary inside the CE needs to be
sufficiently large. This favors systems where both stars are evolved
and have dense and well defined cores (or are only what is left of
their original core, like WDs). 
On top of this, the separation of the
binary inside the CE needs to correspond to GW frequencies roughly
$10^{-4}\,\mathrm{Hz}\lesssim f_\mathrm{GW}\lesssim 0.1\,\mathrm{Hz}$
(gray band in Fig.~\ref{fig:CE_sep}), bound at the low end by the
population of unresolved Galactic WD and at the high end by the
instrument response function \citep[e.g.,][]{robson:19}. The existence
of verification WD binaries in the LISA band, which are likely to have
formed through a CE event, suggests that CE binaries crossing the LISA
band may exist.

To illustrate that some binaries will indeed cross the LISA band
during a CE we use a population-synthesis approach (see also
Sec.~\ref{sec:rate}). In rapid population synthesis calculations, CE
is typically treated as an instantaneous event. This is usually a
sufficient approximation since the duration of a CE is at most of
order of the thermal timescale (depending on the mass, luminosity and
radius of the CE this ranges from several hundreds to several thousand
years), that is much shorter than the entire evolution of a binary
system. To relate the pre-CE initial conditions to the estimated
post-CE configuration, conservation of energy is usually used
\citep[$\alpha\lambda$-algorithms, see e.g.,][]{webbink:84,
  demarco:11}, although algorithms considering conservation of angular
momentum also exist \citep[e.g.,][]{nelemans:00}.

Figure~\ref{fig:CE_sep} shows the post-CE separation
($a_\mathrm{post-CE}$) as a function of the mass that would be
contributing to the GW emission ($M_\mathrm{GW\ emitting}$). In a
rapid population synthesis calculation, each
binary system would drop instantaneously from its
pre-CE separation to the post-CE separation
(as indicated by the vertical arrows). For
illustrative purposes, we assume fully efficient use of the orbital
energy to eject the envelope and no extra energy source (i.e.,
$\alpha=1$), and $\lambda=0.3$ as a typical value for the binding
energy parameter of giant donors
\citep[e.g.,][]{demarco:11}. Increasing either value shifts the curves
to larger post-CE separations.

In Fig.~\ref{fig:CE_sep} we assume $M_\mathrm{GW\ emitting}$ to be the
(helium) core mass of the donor star at the end of the main sequence
plus $0.3\,M_\odot$ assumed to be the mass of a WD companion. We chose
this value to reflect the highest rate of initiation of Galactic CE
events from the simulations described in Sec.~\ref{sec:rate}, which is
expected for systems with $M_\mathrm{core}\simeq 0.5\,M_\odot$ and
$M_\mathrm{WD}\simeq 0.3\,M_\odot$. Although the combination of core
masses with this particular WD mass might not be realistic throughout
the range shown, we have also performed the same calculation assuming
a more massive companion of $1.4\,M_\odot$ (mimicking a Chandrasekhar
mass WD or a neutron star). The resulting curves are shown in
Fig.~\ref{fig:CE_sep_1.4} in Appendix \ref{sec: CE-sep-massive-comp},
each curve moves up to larger post-CE separations, but most curves
still intersect the LISA band (gray area). \referee{For large
  core-radii the final dynamical phase (e. in Fig.~\ref{fig:cartoonCE}
  above) might start at frequencies in the LISA band (hatched area in
  Fig.~\ref{fig:CE_sep}). The systems ending at separation smaller
  than the He core radii could result in mergers rather than
  successful common envelope ejections.  Whether this leads to a
  detectable GW signal or not requires further investigation.}

We use the solar metallicity models from \cite{pols:98} to obtain the
(helium) core mass of the donor from their total masses. Our core masses
are a lower bound since we neglect the core-growth due to the ashes of
(hydrogen) shell burning.  The gray shaded area indicates core
separations corresponding to a GW signal of frequency
$10^{-4}\,\mathrm{Hz}\lesssim f_\mathrm{GW}\lesssim 0.1\,\mathrm{Hz}$,
roughly within the LISA band. Binaries crossing this gray area or
ending within it (e.g., verification WD binaries) are potential
candidates for GW detection of the CE evolution with LISA, depending
on the duration of the CE event. The duration cannot be obtained from
a population synthesis approach, it either needs to be modeled with
challenging multi-dimensional hydrodynamic simulations or, as we will
argue below, it might be directly constrained with observation of the
GW signal.

We return to a full population synthesis calculation of the rate of events in
Sec.~\ref{sec:rate}, after discussing the observability of GW from a
CE in Sec.~\ref{sec:detection}.

\section{Detectability of common envelope in LISA}
\label{sec:detection}

\begin{figure}
  \centering
  \includegraphics[width=0.5\textwidth]{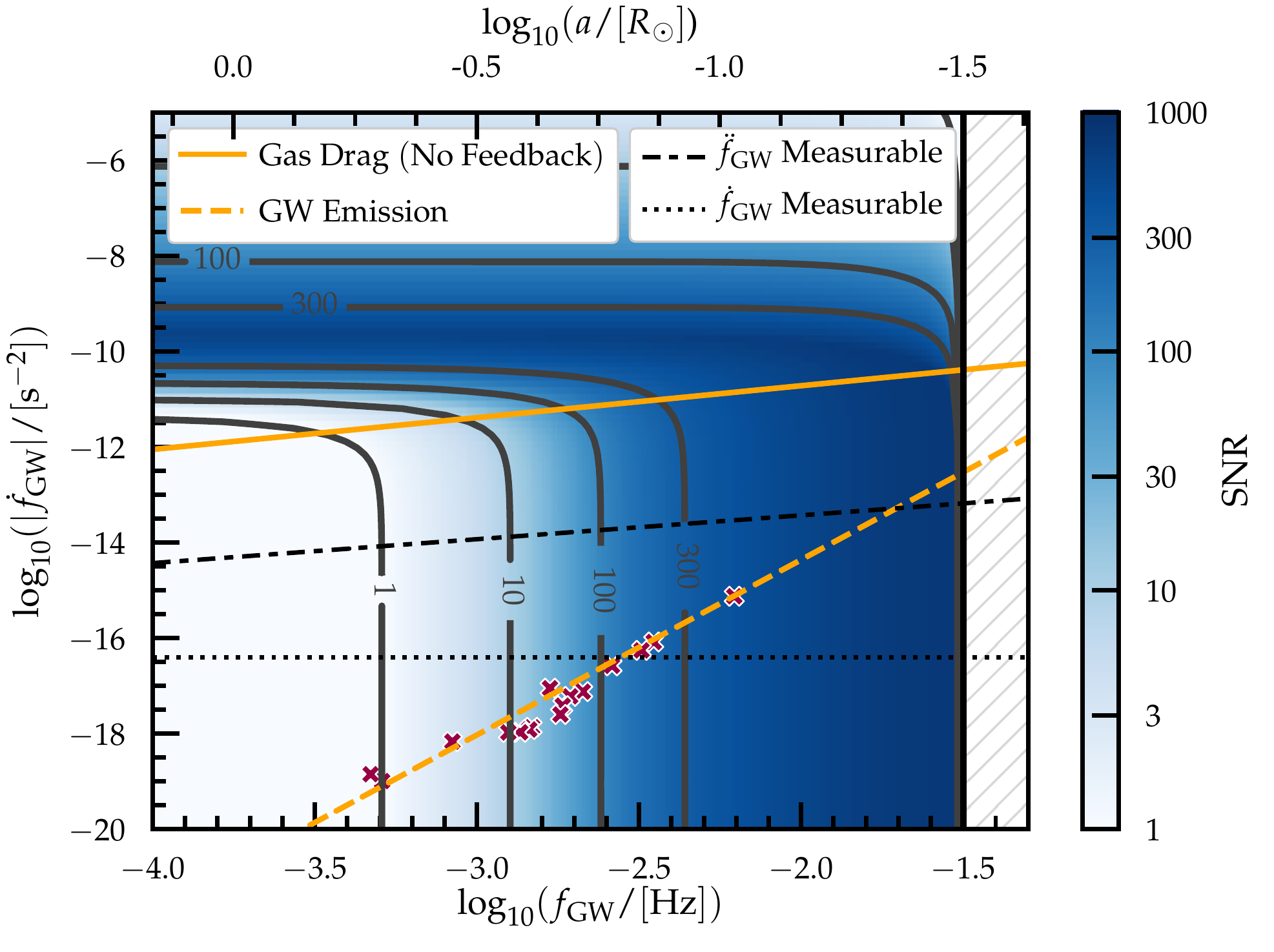}
  \caption{LISA signal-to-noise ratio of a
    \referee{$M_\mathrm{core}=0.5\,M_\odot$ and
      $M_\mathrm{WD}=0.3\,M_\odot$} binary as a function of its
    initial observed GW frequency $f_\mathrm{GW}$ and frequency
    derivative $\dot f_\mathrm{GW}$ \referee{(blue shading and black
      solid lines). For the duration of the observations, we
        assume the shortest between 5\,years or the time to merger at
        constant $\dot f_\mathrm{GW}$, and adopt 3\,kpc as the source
        distance}. The top axis shows the orbital separation
      corresponding to $f_\mathrm{GW}$ for the masses
      considered. Although the exact $f_\mathrm{GW}$ and
      $\dot f_\mathrm{GW}$ values for a binary undergoing CE are not
      known, the dashed and solid orange lines illustrate lower and
      upper bounds on the expected values of $\dot f$ due to CE
      evolution. The former assumes pure GW emission (see
      Appendix~\ref{sec:snr}), and the latter assumes pure gas drag
      with no feedback (see Appendix~\ref{sec:gas_drag}). Finally,
      dotted and dot-dashed black lines mark different regimes in
      which we may or may not expect to successfully
      \textit{distinguish} a CE event from other sources of
      gravitational radiation (see Sec.~\ref{sec:bias}). Frequency
      derivatives above the dotted line are likely detectable after
      five years of observation, and systems above the dot-dashed line
      are further expected to have measurable second derivatives
      $\ddot f_{\rm GW}$, enabling measurement of braking
      indices. Systems lying below these thresholds, meanwhile, are
      most likely to suffer from stealth bias or
      mis-identification. The red crosses indicate known verification
      WD binary.}
  \label{fig:snrs}
\end{figure}

We now explore in what regimes we expect LISA to successfully detect
GW signals from binaries undergoing CE.
Since the details of CE are not yet fully understood, we do
not know the precise frequency evolution of a GW
source undergoing CE.  Instead, in Fig.~\ref{fig:snrs} we
agnostically quantify
the expected signal-to-noise ratios (SNRs) with which LISA will observe
a $M_\mathrm{GW\ emitting}=0.5+0.3\,M_\odot$ binary with an initial GW frequency $\fGW$
and frequency derivatives $\dotfGW$ at a distance of 3\,kpc.
We chose
a value for the typical mass of the donor star core ($0.5\,M_\odot$) and the WD companion
($0.3\,M_\odot$) from our population synthesis (see Sec.~\ref{sec:rate}). Expected SNRs have been averaged over possible sky locations and binary inclinations.
Throughout, we assume the
standard quadrupole radiation such that $\fGW = 2f_\mathrm{orb}$, where
$f_\mathrm{orb}$ is the (quasi-) Keplerian orbital frequency of the CE
binary. We further assume the observation to persist for five years\footnote{The SNR is proportional to the
square root of the observing time.}, or until the given
binary's frequency reaches \referee{$\fGW=10^{-1.5}\,{\rm Hz}$}.
Beyond this frequency, the He core would fill its Roche lobe
within the CE, initiating the dynamical phase that leads to a prompt
merger (bottom right-hand side of Fig.~\ref{fig:cartoonCE}) and
termination of the GW signal.
Additional details of the SNR calculation are given in Appendix~\ref{sec:snr}.

Although the true rate $\dot f_\mathrm{GW}$ at which a binary's GW frequency
evolves under CE is not known, we can nevertheless place sensible
\textit{bounds} on frequency evolution of a CE binary.  In
Fig.~\ref{fig:snrs}, the orange dashed line shows the frequency
derivative $\dot f_\mathrm{GW}(f_\mathrm{GW})$ due solely to the emission of GWs.
For a given total mass, this is a strict lower bound on the evolution rate of a binary
undergoing CE. Meanwhile, the orange solid line shows the value of
$\dot f_\mathrm{GW}(f_\mathrm{GW})$ expected in the case of pure gas
drag, neglecting the reaction of the envelope and assuming
$\rho=10^{-6}\,\mathrm{g\ cm^{-3}}$,
$T=\referee{9\times}10^5\,\mathrm{K}$ \referee{to represent} the
density and temperature \referee{deep inside} the shared CE (see
Appendix~\ref{sec:gas_drag}). This provides a strict upper-limit,
since in nature we expect the envelope to expand due to the energy injected into the envelope by the
inspiral, lowering the gas drag on the binary inside the CE.
Generally, we expect CE binaries to lie between these two bounds.
Under these constraints, we see that a systems occupying a large
portion of the remaining $f_\mathrm{GW}-\dot f_\mathrm{GW}$ phase
space are detectable with LISA.  In particular, systems whose
gravitational wave frequency is $f\referee{_\mathrm{GW}} \gtrsim 10^{-3}\,{\rm Hz}$
(corresponding to orbital separations less than
$\sim{}0.3\,R_\odot$ \referee{for our fiducial masses})
yield ${\rm SNR}\gtrsim 10$.

Although galactic binaries undergoing self-regulated CE evolution may be detectable by LISA, they are not necessary \textit{identifiable} as such.  Without additional information or careful consideration, we run the risk of misidentifying such a system as a purely GW-driven compact binary, incorrectly attributing an anomalously high $\dot
f_\mathrm{GW}$ to large component masses rather than accelerated evolution due to CE.  This issue of identification will be discussed further in Sec.~\ref{sec:bias} below.

\section{Estimate of the number of Galactic common-envelope binaries}
\label{sec:rate}

To determine the number of CE binaries in the Galaxy that LISA (and
future GW observatories) might be able to detect, we use the
population synthesis code
COSMIC\footnote{\href{https://cosmic-popsynth.github.io/}{https://cosmic-popsynth.github.io/}. Our
  input files and results are available at
  \url{https://zenodo.org/record/4490011}.} \referee{version 3.2}
\citep{breivik:20}. As in all rapid binary population synthesis codes,
COSMIC treats CE as an instantaneous event, therefore, we obtain from
COSMIC the rate at which CE event are initiated ($R_\mathrm{CE,init}$)
by scaling the number of binaries initiating a CE with the total mass
of the population and multiplying by a star formation rate (SFR) of
$3.5\,M_\odot\ \mathrm{yr^{-1}}$ \cite[commonly used for Milky-way
equivalent galaxies in population synthesis calculations of GW
sources, e.g.,][]{dominik:12}. The value of $R_\mathrm{CE,init}$ can
be rescaled for an arbitrary SFR, which anyway does not constitute the
dominant uncertainty in the determination of the number of Galactic
sources.

Population synthesis calculations assume a CE event occurs if one of the
following condition is met: (i) both stars simultaneously fill their
Roche lobe (e.g., for initially equal mass ratio systems), (ii) one of
the stellar radii exceeds the periastron point -- for eccentric
binaries, or (iii) when the mass transfer is deemed to be
unstable. Case (iii) is most common. Mass transfer stability
is decided in COSMIC using critical mass ratio values $q_c$
(\citealt{breivik:20}, see also \citealt{hurley:02}, \citealt{pavlovskii:17},
\citealt{vigna-gomez:20}). Whenever the mass ratio between the donor
star and the accretor is larger than a given threshold
$M_\mathrm{donor}/M_\mathrm{accretor}\geq q_c$, mass transfer is
defined to be unstable and a CE is initiated.
Different $q_c$ values can be used for each
evolutionary phase of the donor star, to reflect the changes in thermal
timescales as stars evolve.

\begin{figure*}[hbtp]
  \centering
  \includegraphics[width=0.49\textwidth]{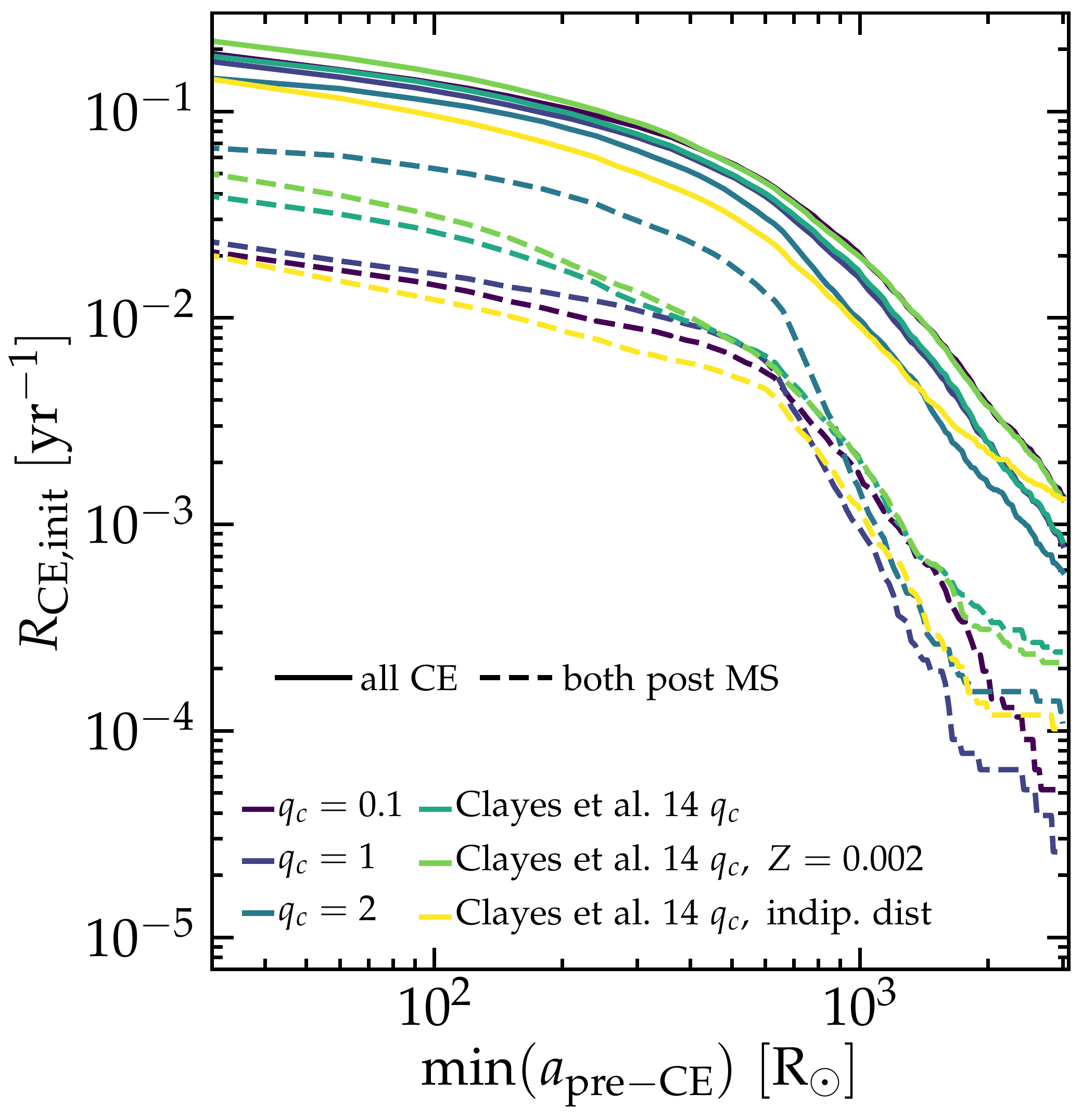}
  \includegraphics[width=0.49\textwidth]{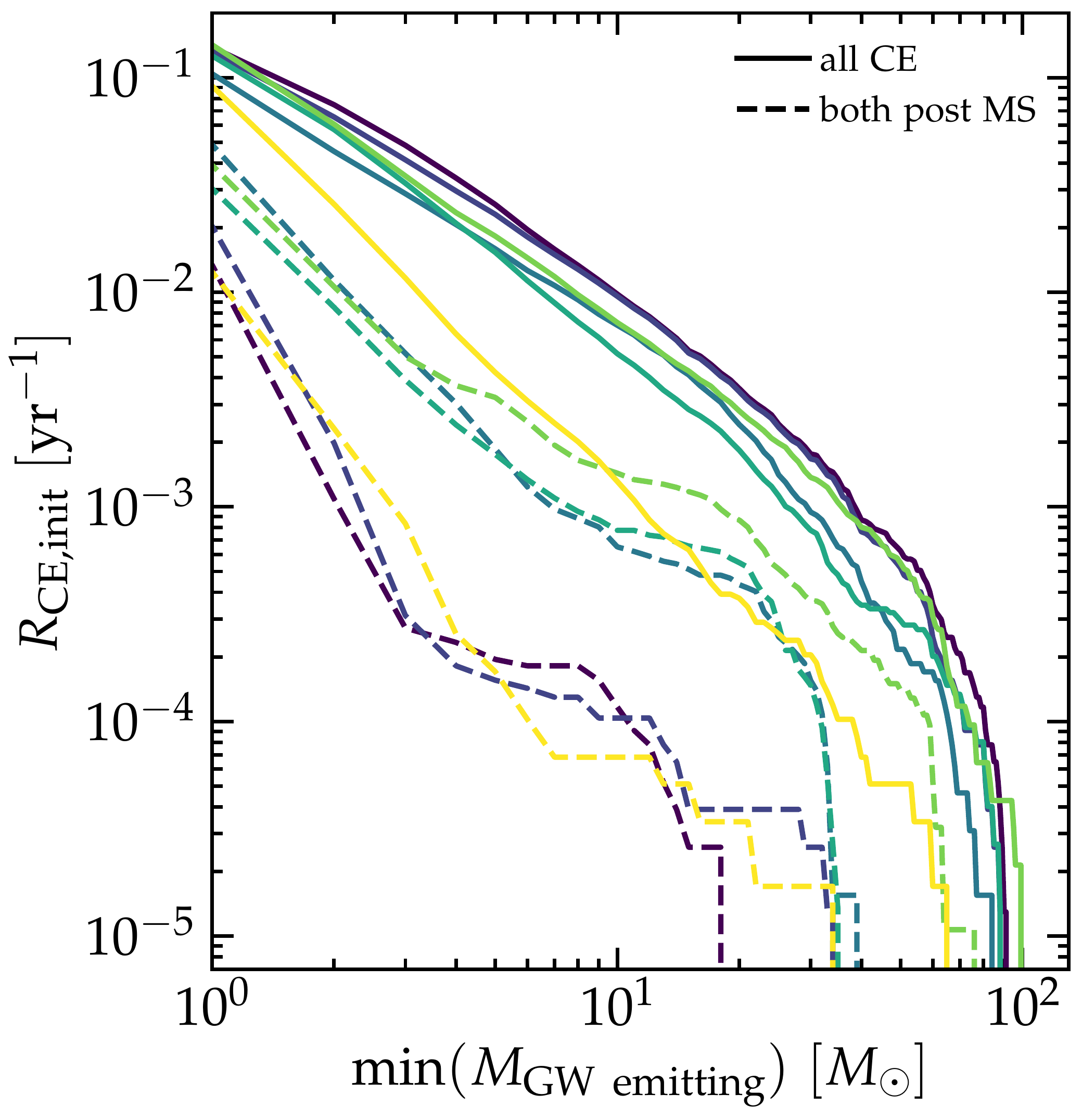}
  \caption{Rate at which CE events are initiated $R_\mathrm{CE,init}$
    as a function of the minimum pre-CE separation (left) and minimum
    total mass emitting GWs (right). The solid lines include all CE
    events, while the dashed line only include CE where both stars are
    beyond the main sequence, or in other words, where two dense
    stellar cores are involved. Different colors correspond to
    different assumptions on the stability of mass transfer ($q_c$,
    see also text) and initial distribution of binaries
    (yellow).}
  \label{fig:rates}
\end{figure*}

In our fiducial run -- labeled as ``Clayes et
al. 14 $q_c$'' in Fig.~\ref{fig:rates}, we assume the
\referee{donor-type dependent} $q_c$ values\footnote{\referee{However, \cite{claeys:14} define the critical mass ratio as $M_\mathrm{accretor}/M_\mathrm{donor}$.}}
\referee{from Tab.~2 of} \cite{claeys:14}, a
metallicity $Z=0.02$, and the correlated initial distributions in
primary mass, mass ratio, period, and eccentricity from
\cite{moe:17}.
We explore several variations of our fiducial run, keeping the same
setup except one of the following:
\begin{itemize}
\item constant $q_c$ for all evolutionary phases of the donor star. We
  consider values from $q_c=0.1$ (more likely to enter CE) to $q_c=2$
  (less likely to enter CE);
\item lower metallicity $Z=0.002$;
\item independent and uncorrelated initial distributions -- with a
  \citealt{kroupa:93} initial mass function for the primary stars, a
  flat mass-ratio distribution, a log-normal period distribution
  \citep[e.g.,][]{duchene:13}, and a thermal eccentricity
  distribution.
\end{itemize}

COSMIC is a Monte Carlo population synthesis code \citep{breivik:20},
and for each population described above, we stop drawing systems when the sampling
uncertainty in the initial distributions is less than $10^{-5}$,
that is drawing more systems has an effect smaller than this
threshold. More specifically, we check the initial primary and
secondary mass distributions in the range [0,80]\,$M_\odot$, and
initial orbital period below 1000\,days.  The number of systems
simulated to reach this threshold ultimately determines the total mass
of the simulated population, which we use to determine
$R_\mathrm{CE,init}$.

These parameter variations do not exhaust the entire range of
uncertainties. For example, the fraction of mass that is accreted
during stable mass transfer can influence the orbital evolution of the
binary, and thus modify the chances of a subsequent reverse CE from
the initially less massive star to the initially more massive
\citep[e.g.,][]{zapartas:17}. However, we expect the impact of other
parameters to be smaller than the impact of varying $q_c$ and the
initial distributions for binaries.

We find a rate of CE initiation regardless of the masses involved of
$R_\mathrm{CE,init}=0.18^{+0.02}_{-0.09}\,\mathrm{yr^{-1}}$ for all CE
events, and $0.06^{+0.03}_{-0.02}
\,\mathrm{yr^{-1}}$ when considering
only CE between two evolved stars (see below). The uncertainties on
$R_\mathrm{CE,init}$ are significantly smaller than the uncertainty on
the duration of the CE (cf.~Sec.~\ref{sec:ce_theory}).

Figure~\ref{fig:rates} shows the rate of CE initiation that we obtain
as a function of the minimum pre-CE separation (left) and the minimum
mass emitting GWs (right), i.e. the rate of events with pre-CE
separation (or GW emitting mass) indicated by the abscissa or
larger.
For example, for our fiducial simulation, about $1.6\times10^{-2}$ CE events are initiated per
year with a pre-CE separation
$a_\mathrm{pre-CE}\geq 10^3\,\mathrm{R_\odot}$.  The solid lines in
Fig.~\ref{fig:rates} show all CE events initiated, while the dashed
lines focus on CE events where both stars are evolved beyond the main
sequence, meaning both have a well defined, dense core. The latter
(dashed lines) represent systems that are more likely to be detectable
in GWs.

\referee{When considering all CE events, higher $q_\mathrm{crit}$
  correspond to lower $R_\mathrm{CE,\ init}$. Conversely, CE with both
  stars post-MS, i.e. with a (semi-) degenerate core, might not be the
  first mass transfer episode in a given binary, and their occurrence
  and rate depend also on what is assumed for previous mass transfer
  phases.} The CE initiation rate is rather flat for pre-CE
separations as large as $a_\mathrm{pre-CE}\lesssim700\,R_\odot$, while
it drops very quickly with increasing $M_\mathrm{GW\ emitting}$ because of the
initial mass function.

To determine the mass emitting GWs in each binary, we assume the
following: if a star (donor or accretor) is on the main sequence (MS),
that is, it lacks a well defined core, we use its total mass in the
mass emitting GWs.  If a star has evolved beyond the MS, only the core
mass adds to the GW emitting mass.  If a star has become a compact
object (WD, neutron star or black hole) again the total mass of the
object adds to the GW emitting mass. There are two edge cases where
our assumptions might overestimate the amount of mass involved in
GW-emission as the total mass of the binary: CE between two MS stars,
and between two compact stars (e.g., a double-degenerate SNIa
progenitor, \citealt{dan:11}). However, this has very little
effect on our conclusions. With our initial period distribution, CE
between two MS stars are relatively rare, and they are unlikely to be
the most interesting target for LISA, since the lack of dense cores
limits the mass involved in the generation of GWs and thus the signal
amplitude. 
Setting instead the total mass emitting GWs to zero
for these systems has a negligible effect on the right panel of
figure~\ref{fig:rates}. CE between two compact stars do not occur with
our fiducial setup.

Assuming the stellar population of the Galaxy is in equilibrium,
an estimate of the current number of Galactic CE can then be obtained with
\begin{equation}
  \label{eq:N_CE}
  N_\mathrm{CE} = R_\mathrm{CE,init}\times \Delta t_\mathrm{CE}  \ \ ,
\end{equation}
where $\Delta t_\mathrm{CE}$ is the typical duration of a CE event,
\referee{which is not known. Moreover,} what matters for GW detection
is not the actual duration $\Delta t_\mathrm{CE}$, but the fraction of
it spent at a separation corresponding to a GW frequency in the
detector bandpass.

Depending on the \referee{evolutionary stage of the CE binary} and the model assumed,
\referee{estimates of the} CE duration $\Delta t_\mathrm{CE}$ can
range from dynamical timescale ($\sim$ days, e.g., \citealt{ginat:20,
  lawsmith:20}) to thermal timescale ($\sim10^5\,\mathrm{years}$), although the radial expansion and
luminosity increase of the envelope during the CE are likely to reduce
the relevant thermal timescale to $10^3-10^4$ years
\citep[e.g.,][]{meyer:79, clayton:17, igoshev:20}. Recent hydrodynamical
simulations suggest a
$\sim$$10$\,years duration \citep[e.g.,][]{fragos:19, chamandy:20}.

\referee{Based on these previous estimates}, assuming a short duration of the CE of
$\sim$\,years, the rate of CE at the low mass end ($R_\mathrm{CE,init}\approx
0.1$\,yr$^{-1}$) is sufficiently large that at least one source might be detectable during a 10-year LISA mission.  Considering only CE with evolved stars (dashed lines), $R_\mathrm{CE,
  init}$ drops to
$\sim0.5\times10^{-2}$\,yr, making the expected number of sources only marginally non-zero.

If the self-regulated phase of the CE instead lasts much longer than
the planned LISA mission, there might be upwards of hundreds of
detectable sources in the Galaxy
($R_\mathrm{CE,init} \approx 0.1$\,yr$^{-1}$ $\times 10^4$ yr
$\simeq 1000$).  Thus, under the most pessimistic assumptions, there
might be only a slim chance of detecting a CE event through LISA,
while under more \referee{optimistic, but not extreme, assumptions}
there could be a significant number of target sources available. We
return on the possible implications of a non-detection in
Sec.~\ref{sec:non-detection}.

\section{Stealth bias}
\label{sec:bias}

There is an ever growing catalog of potential systems that LISA might
be sensitive to, both galactic and extragalactic.  Examples include
exoplanets~\citep[e.g.,][]{2019NatAs...3..858T}, WD binaries in
various stages of their evolution \citep[e.g.,][]{kupfer:20, liu:21}, and
binary black holes (BBH) in vacuum or embedded in gas
\citep[e.g.,][]{chenX:20}. Our study suggests that CE events are also
potentially detectable by LISA, raising the question: how can we
distinguish GWs from a CE from these other possibly confounding
sources?  In this section we discuss GW-based ways that could be used
to conclude that a detected signal is a CE event and the expected bias
if we fail to do so. We discuss further EM-based ways of resolving
this stealth bias in Sec.~\ref{sec:EM_counterparts}.

Depending on the exact frequency evolution of the CE event, there may
exist an observational degeneracy between the CE inspiral and purely GW-driven
binaries of higher mass hosting black holes and/or neutron stars.
In standard GW analysis that assumes a binary in vacuum is described by
general relativity (GR), a measurement of a binary's $f_{\rm GW}$ and $\dot f_{\rm GW}$ can be used to estimate its mass and distance.
If a particular binary is actually undergoing CE, the assumption of vacuum breaks down and
the resulting estimates will be biased compared to their true values \citep[see also][ for an extended discussion]{chenX:20}.

\begin{figure}[htbp]
  \centering
  \includegraphics[width=0.5\textwidth]{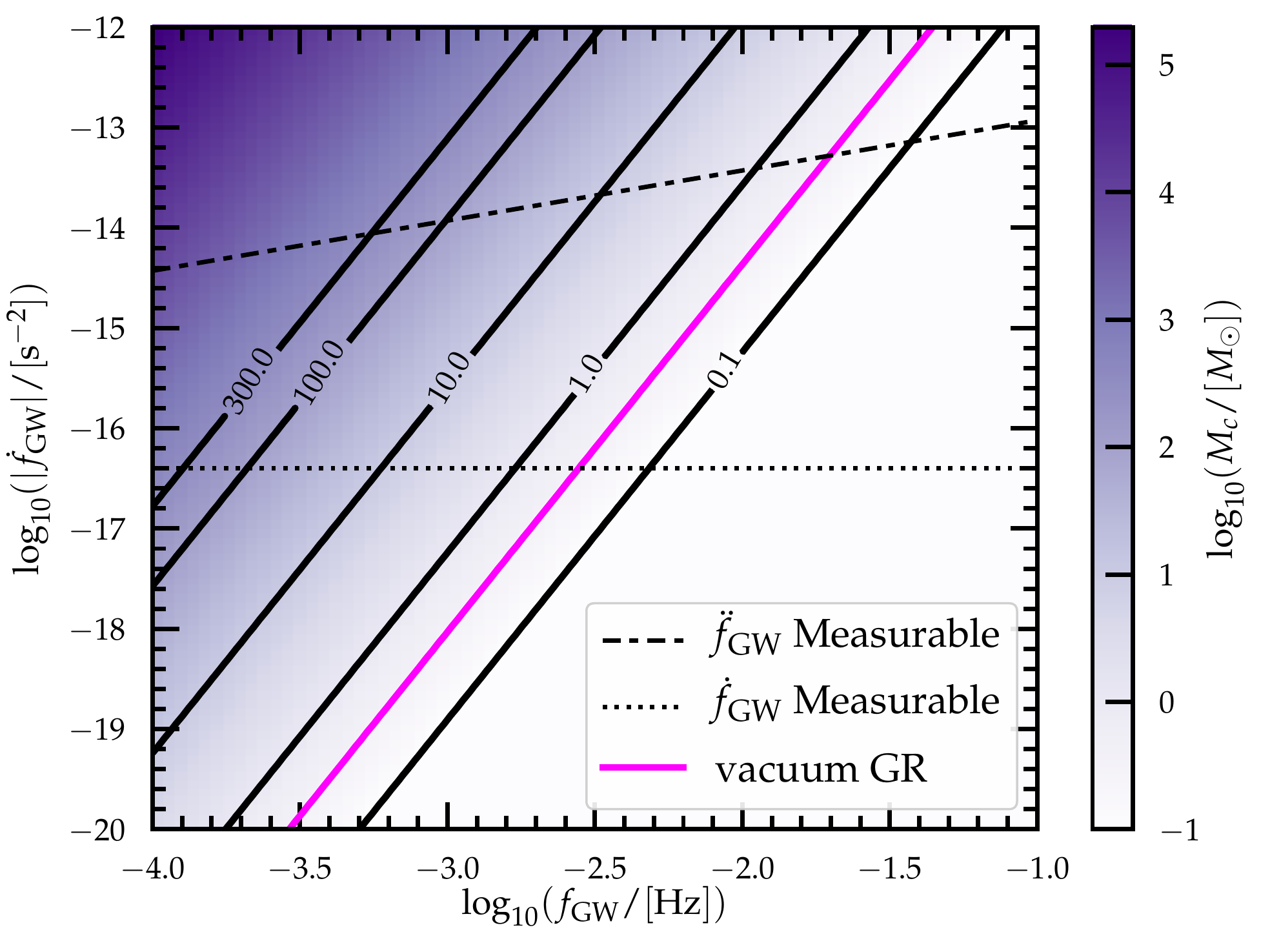}
  \includegraphics[width=0.5\textwidth]{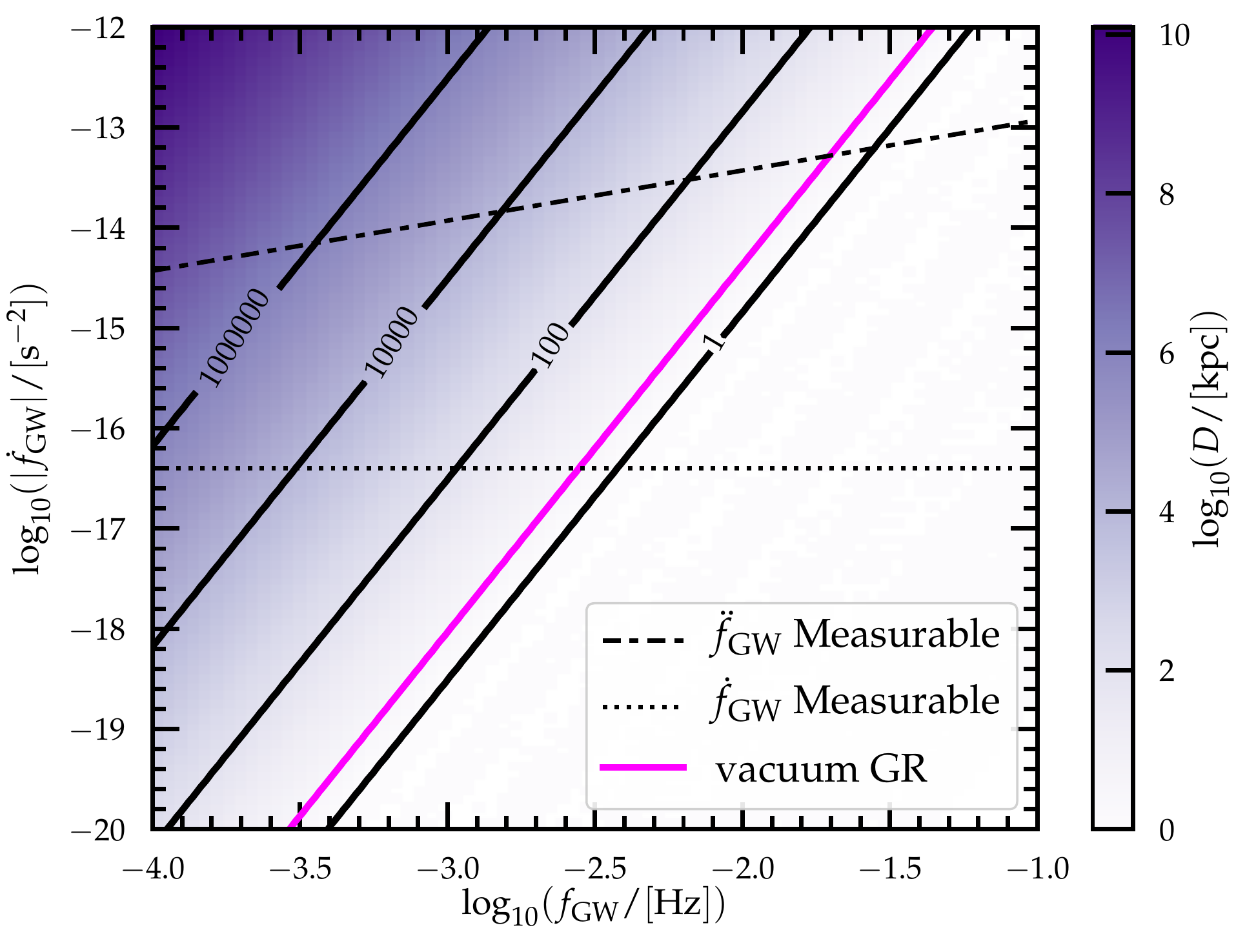}
  \caption{ Inferred chirp mass
    (top) and distance (bottom) for each measured $f_\mathrm{GW}$ and $\dot{f}_\mathrm{GW}$ for a
    system with
    $M_\mathrm{core}=0.5\,M_{\odot}, M_\mathrm{WD}=0.3\,M_{\odot}$ at
    $3$\,kpc.  Similar to Figure \ref{fig:snrs}, we use a a dot-dashed
    line to show the border above which $\ddot{f}_\mathrm{GW}$ can be measured and
    the nature of the binary can be determined through the braking
    index measurement (see Sec.~\ref{sec:braking_index}).  Below that
    line there is a possible stealth bias, which these two plots
    quantify.  Above the dotted line, frequency derivatives are likely
    detectable after five years of observation. Systems lying below
    the dotted line would be observed as stationary sources.  The
    magenta line corresponds to the correct value of the chirp mass
    and distance, namely the GR relation between $f_\mathrm{GW}$ and $\dot f_\mathrm{GW}$ for
    these masses. Hence the bias
    vanishes.}
  \label{fig:bias}
\end{figure}

This bias is quantified in Fig.~\ref{fig:bias}, which shows the chirp mass
\begin{equation}
  \label{eq:chirp_mass}
  M_c = \frac{(M_1M_2)^{3/5}}{(M_1+M_2)^{1/5}} \ \ ,
\end{equation}
(color in top) and distance ($D$, color in bottom) that we would
\textit{incorrectly} assign to a $0.5+0.3\,M_\odot$ binary undergoing
CE at $3$\,kpc, if we have measured only its gravitational-wave
frequency and frequency derivative (and the signal amplitude). We note
that for our example binary, the correct value of the chirp mass would
be $M_c=\referee{0.34}\,M_\odot$ (pink line in the top panel). Depending on the
exact values of ($f_\mathrm{GW}$, $\dot f_\mathrm{GW}$), we find the
possibility of large biases in the inferred parameters, some of them
being consistent with BBHs at cosmological distances.

We identify three possible means of breaking this degeneracy between BBHs and CE events:
\begin{itemize}
\item[(i)] Increased observation time could result in a measurement of
  $\ddot f_\mathrm{GW}$ (and higher derivatives) in addition to
  $f_\mathrm{GW}$ and $\dot f_\mathrm{GW}$, enabling a measurement
  of the binary's \emph{braking index}. This scenario is further
  discussed in Sec.~\ref{sec:braking_index}.
\item[(ii)] The (incorrectly) inferred chirp mass may be high enough that the system is
  consistent with a BBH.
  In this case, one would expect to continue to see the signal evolve,
  and eventually observe the merger in the LIGO band.
  Failure to observe the final merger, or a signal that is different from a BBH merger signal in
  vacuum GR would indicate that the binary may have been
  misidentified.
\item[(iii)] The simultaneous measurement of an apparent high chirp
  mass and a low distance might indicate the presence of a BBH close
  to Earth, potentially even in the Galaxy for certain combinations of
  $f_\mathrm{GW}$ and $\dot f_\mathrm{GW}$. Such a system might be
  inconsistent with independent information about the BBH inspiral and
  merger rates obtained from ground-based detectors. This possibility
  is further discussed in Sec.~\ref{sec:rate_bias}.
\end{itemize}
If these outcomes fail, the possibility of a \emph{stealth} bias on
  the system parameters remains present, it may be impossible to conclusively identify the true nature of the given binary.

\subsection{Measuring the braking index}
\label{sec:braking_index}

Stellar binaries evolving purely via GW radiation have
GW frequencies that evolve in time as $\dot f_\mathrm{GW} \propto f_\mathrm{GW}^{11/3}$.
In general, given some source evolving as $\dot f \propto f^n$, a
measurement of the braking index $n$ given by
    \begin{equation}
    n = \frac{f \ddot f}{\dot f^2} \ \ ,\label{n-def}
    \end{equation}
is possible if the \textit{second} time derivative $\ddot f$ is
measurable. We emphasize the analogy with the pulsar braking index,
where $f$ is the pulsar spin frequency rather than the frequency of GW
radiated by the CE binary.

If LISA can successfully measure $f_\mathrm{GW}$,
$\dot f_\mathrm{GW}$, and $\ddot f_\mathrm{GW}$ for a given GW source,
inference on the source's braking index  offers a means of
identifying the physical processes driving its evolution.
In particular, we can seek to verify if $n$ is consistent with
$n_\mathrm{GW}=\frac{11}{3}$, as expected for pure gravitational
radiation in vacuum, or inconsistent with $n_\mathrm{GW}$, possibly
indicating a binary in the midst of CE or subject to other
environmental influences \citep[e.g.,][]{fedrow:17}.

While LISA is expected to measure $\dot f_\mathrm{GW}$ for a considerable number
of binaries, $\ddot f_\mathrm{GW}$ is likely measurable only in a much smaller
subset of cases. Although determining when $\dot f_\mathrm{GW}$ and $\ddot f_\mathrm{GW}$ are
measurable requires expensive
simulations~\citep{Seto:2002dz,Littenberg:2020bxy},
we can make an order-of-magnitude estimate by considering how LISA's frequency resolution scales with time.
As LISA accumulates
more data, its frequency resolution scales with the observing time $T$
as $\Delta f_\mathrm{GW} \approx 1/T$. Accordingly, we assume that $\dot f_\mathrm{GW}$ and
$\ddot f_\mathrm{GW}$ are measurable when they cause a binary's frequency to
change enough such that it moves from one frequency bin to another
over the course of LISA's lifetime. Then, \referee{lower limits on the} minimum detectable
$\dot f_\mathrm{GW}$ and $\ddot f_\mathrm{GW}$ are
\begin{align}
    \dot f_{\rm GW, min} T &= \Delta f_{\rm GW} \quad \Rightarrow \quad \dot f_{\rm GW, min} \approx \frac{1}{T^2} \  \ ,\\
    \frac{1}{2}\ddot f_{\rm GW, min} T^2 & = \Delta f_{\rm GW} \quad \Rightarrow \quad \ddot f_{\rm GW, min}\approx\frac{2}{T^3} \ \ .
    \label{eq:fddot}
\end{align}
\referee{We refer the readers to \cite{takahashi:02} and
  \cite{robson:18} for more accurate estimates accounting for the SNR
  dependence.}

On the
$\dot f_\mathrm{GW}-f_\mathrm{GW}$ plane, the first condition is a
straight line depicted by the black dotted horizontal line in
Fig.~\ref{fig:snrs} and \ref{fig:bias}.
We can also identify the region of this plane in which $\ddot f_{\rm GW}$ is measurable for a system evolving in vacuum GR.
Using the definition of braking index
(Eq.~\ref{n-def}) to express $\dot f_\mathrm{GW}$ as a function of
$f_\mathrm{GW}$ and $\ddot f_\mathrm{GW}$, a binary evolving purely via GW emission with the minimum detectable $\ddot f_{\rm GW}$ (Eq.~\ref{eq:fddot}) has a frequency derivative
\begin{equation}
\dot f^2_{\rm GW} = \frac{3}{11}\ddot f_{\rm GW, min} f_{\rm GW} \ \ .
\label{eq:fddot_line}
\end{equation}
This relation is plotted in Figs.~\ref{fig:snrs} and \ref{fig:bias}
with black dot-dashed line.  Consider a system lying above this line.
Such a system \emph{should} have a measurable $\ddot f_{\rm GW}$ (and
hence a measurable braking index) if it evolves via GW emission.
Therefore, if this system's $\ddot f_{\rm GW}$ or $n$ is not measured,
or is measured to be different than the GR prediction, then the
non-vacuum or non-GR nature of the binary can be inferred.  Next,
we consider systems lying below the line defined by
Eq.~\eqref{eq:fddot_line}. For such systems, GW evolution alone cannot
lead to a detectable $\ddot f_{\rm GW}$.  If $\ddot f_{\rm GW}$ is
nevertheless measured, one would infer that the binary does not evolve
solely under vacuum GR effects. If $\ddot f_{\rm GW}$ is not measured, then it is not possible to determine the nature of the
binary.

\subsection{Distinction from other sources bases on rates}
\label{sec:rate_bias}

One potential way to infer the nature of sources that cannot be
unambiguously identified using only their GW signature is through
their rate of occurrence. For example, if the rate of CE detections in
LISA is higher than the uncertainty in the rate of BBH mergers, one
can statistically filter out the over-abundance of BBH detections in
LISA when comparing to the BBH merger rate from ground-based
detectors.

As of LIGO/Virgo O3a, the rate of BBH mergers
\citep[29.9$^{+14.9}_{-8.6}$Gpc$^{-3}$yr$^{-1}$,][]{GWTC2,GWTC2pop}
is relatively poorly constrained due to the small sample size of
BBH events and the dependence on the assumed BH mass distribution.
However, when LISA will become operational \citep[around the year 2035,][]{LISAproposal2017}
the BBH catalog size from ground based detectors will have increased to
approximately $10^{4.5}$--$10^5$ BBH detections per year \citep[e.g.][]{baibhav:2019}.
This means that the constraints on the BBH merger rate will have improved
drastically.

The LISA mission will result in an independent measurement of the rate of BBH mergers,
 which will be based on the GW signals from wide (in-spiraling) BBH systems.
If the BBH merger rate as inferred from LISA is significantly higher than the,
by then, well constrained BBH merger rate from ground based detectors,
one could argue that this overabundance could originate from, for example, CE-events.

Moreover, the BBH merger rate from ground based detectors will
be sufficiently constrained to allow for a split based on event parameters such as BH mass and spin.
This will enable us to constrain, for example, the rate of massive BHs ($M_{\rm{BH}} \sim 40$ M$_{\odot}$) in the local Universe.
CE-events might appear as more massive BBH events (see Fig.~\ref{fig:bias})

Whether or not the discrepancy between the BBH merger rates inferred
from LISA with respect to the ground based detectors will be
significant, will depend both on the size of the discrepancy and the
uncertainty in the rate measurement from LISA.  If we assume that such
an overabundance in the rate comes from CE-events, then the size of
the discrepancy is directly related to the rate of CE events.  Based
on our discussion in Sec.~\ref{sec:rate} we expect the number of CE
events detectable by LISA, to lie between 0.1 and 1000. Although this
estimate spans several orders of magnitude, we will be able to
constrain the upper limit of this estimate (corresponding to long
lasting CE-events with $\Delta t_\mathrm{CE} \propto 10^4$yr), given
the uncertainty in the stellar mass BBH-merger rate in LISA.


\section{Sky localization and electromagnetic counterparts}
\label{sec:EM_counterparts}

EM transients could provide interesting triggers for GW searches, and
in this case the probability of misinterpreting the signal might be
lower. In fact, the self-regulated inspiral phase is expected to
follow the dynamical plunge-in phase (step (\emph{c}) in
Fig.~\ref{fig:cartoonCE}), which might cause optical \citep[e.g.,
LRNe][]{Soker:2003,Kulkarni:2007,ivanova:13b, Pastorello:2019} and
infrared (IR) transients \citep[e.g., SPRITES][]{Kasliwal:2017}.

An example of what a possible EM precursor could look like is V1309Sco
\citep{tylenda:11}, and given that the initial parameters of the
binary stars are known fairly well, it would be interesting to monitor
this object for possible GWs once LISA is operational. Concerning the
characterization and discovery of EM precursors, the spectral
evolution and late-time observations of V1309Sco suggest that LRNe
could produce large amounts of dust and be particularly bright in the
infrared \citep[IR,][]{pejcha:17, Metzger:2017, Iaconi:2020,
  Nicholls:2013, Jencson:2019, Blagorodnova:2020}.  Therefore, it is
conceivable that some of the transients associated with CE events
might only be observable in the IR.  The SPitzer InfraRed Intensive
Transients Survey (SPIRITS) detected unusual infrared transients with
no optical counterparts \citep{Kasliwal:2017}. These transients, dubbed
eSPecially Red Intermediate-luminosity Transient Events (SPRITEs), are
in the infrared luminosity gap between novae and supernovae, and could
be associated with stellar mergers and CE events.

Another way to identify possible interesting targets where the CE
event might have been initiated too long ago to detect its beginning
is through peculiar circumstellar environments. An example could be
TYC 2597-735-1 which was interpreted as a merger product based on its
``blue-ring nebula'' but still exhibits variability \citep[][although
its variability is not necessarily related to a still ongoing
merger]{hoadley:20}.

Conversely, in the case of a GW detection, the
interpretation of the signal might be helped by EM followup. This
might be necessary to distinguish a CE signal from other potential mHz
GW sources (e.g., AM CVns, sdO+WD binaries). As mentioned before, the
most promising GW signal for LISA does not come from the dynamical
phase of a CE associated to the brightest EM transients (e.g.,
LRNe). Instead, the long-lasting self-regulated inspiral produces a
possibly slowly-evolving GW source. The LISA sky localization is
strongly dependent on the signal and its SNR, for a monochromatic
source typical uncertainty ellipses are of order tens of degrees
squared \citep{cutler:98}.

During the self-regulated phase, the CE binary might
look like a red giant with the following possible features.
The surrounding circumstellar material might be peculiar:
a CE is initiated by an unstable phase of RLOF (step
\emph{a} in Fig.~\ref{fig:cartoonCE}), which can
be non-conservative and spill over in the surroundings of the system
\citep[e.g.,][]{pejcha:17, macleod:20}. One-dimensional CE simulations
suggest that periodic outbursts of mass ejection during the
self-regulated phase are possible \citep[e.g.,][]{clayton:17}. Thus,
a CE binary might appear as a red giant with an excess of circumstellar
material and peculiar time-variability.

\section{Discussion}
\label{sec:discussion}

\subsection{Comparison to other common envelope rate estimates}
\label{sec:other_rates}

Assuming a steady-state stellar population in the Galaxy, the number
of CE binaries viable as GW sources depends on the initiation rate
$R_\mathrm{CE,init}$ and the time duration when the system might be
detectable. While presently the latter needs to be estimated from numerical
simulations (although, see also \citealt{igoshev:20}), the former can
be constrained using observations of EM transients
associated with CE events. 

The most relevant EM transient to constrain the rate of CE are LRNe,
which have been conclusively linked to a CE event via
observations of the period decay preceding the outburst event in
V1309Sco \citep{tylenda:11}. \cite{kochanek:14} estimate the Galactic
rate of these events with $V$ ($I$) band magnitude brighter than $-3$
($-4$) to be $\sim0.5\,(0.3)\,\mathrm{yr^{-1}}$, in reasonable agreement
with our $R_\mathrm{CE,init}=0.18^{+0.02}_{-0.09}\,\mathrm{yr^{-1}}$
(cf.~Sec.~\ref{sec:rate}). Their estimate was based on the then
available sample of 4 events in 25 years and correction for
observational biases. \cite{kochanek:14} also noted the good
agreement with population synthesis results obtained with the
\texttt{StarTrack} code. More recently, \cite{howitt:20} used the
COMPAS population synthesis code to suggest a Galactic rate of
LRNe of 0.2$\,\mathrm{yr^{-1}}$, which is in agreement with
our estimates obtained with COSMIC.

\subsection{Consequences of non-detection}
\label{sec:non-detection}

Even a non-detection of GWs from a CE event throughout the LISA
mission might provide unique and direct constraints on CE
evolution. In this case, regardless of the rate of CE initiation, the
first possibility is that the self-regulated thermal-timescale phase
happens at too wide separations, corresponding to
$f_\mathrm{GW}\lesssim 10^{-4}$\,Hz below the LISA band pass. For our
representative CE binary with $M_\mathrm{core}=0.5\,M_\odot$ and
$M_2=0.3\,M_\odot$, this would mean
$a_\mathrm{post-plunge}\gtrsim1.5\,R_\odot$, although from
Fig.~\ref{fig:snrs} we expect to build up significant SNR only
for $a_\mathrm{post-plunge}\lesssim 0.5\,R_\odot$.

Alternatively, direct
constraints on the duration in the LISA band ($\Delta t_\mathrm{CE}$)
can be derived with some informed assumptions on the rate of
initiation ($R_\mathrm{CE,init}$).  Considering all CE events, the
latter can be estimated with the observed rate of LRNe at
$\approx 0.1$\,yr$^{-1}$ (consistent with our population synthesis
calculations, see also Sec.~\ref{sec:other_rates}). Restricting to CE
between two evolved stars with dense cores, our simulations suggest a
decrease of about a factor of $\sim$3 (cf.~Sec.~\ref{sec:rate}).
Therefore, despite the many uncertainties in binary mass transfer
stability and CE evolution, both EM transients observations and
population synthesis estimates suggest a value of about one CE
initiation every decade in the Galaxy. Assuming most of these event
will cross the LISA band at some point (cf.~Fig.~\ref{fig:CE_sep}) and
a mission duration of $T\approx 10$\,yr, a non-detection would be at
odds with models for which $\Delta t_\mathrm{CE}\gg 1$\,yr, which we
predict might result in a significant number of detectable
sources. The longer the time $T$ during which a GW signal might be
observed is, the less dependent on uncertainties in
$R_\mathrm{CE, init}$ and more stringent this constrain will be.
\newpage
\subsection{Further caveats}

There are several effects which we have not explored in detail in this
study which could impact the LISA detectability of a CE event.  The
dynamical plunge-in could make the orbit inside the CE eccentric,
modifying the GW emission in two ways: make it more ``bursty'', with
stronger emission at each pericenter passage, and make it more
``directional''. If the rate of CE events is on the optimistic side,
LISA can still expect to detect eccentric CE events, and will unveil a
great deal about the physics of the dynamical plunge-in.

We have also neglected triple stellar systems. Triples can enhance the
rate of mass transfer in binaries, either because the third star fills
its Roche lobe, or because of the Kozai-Lidov oscillations inducing
high eccentricity in the inner binary \citep[e.g.,][]{toonen:20}. This
can in principle enhance the rate of CE and produce events which are
qualitatively different in EM and GWs, see for example
\cite{glanz:21}. Hence, our most pessimistic rate calculations may be
regarded as a lower limit.

\section{Summary and conclusions}
\label{sec:conclusions}

Common-envelope evolution remains one of the largest uncertainties in stellar
physics. Here, we have investigated the possibility of detecting
gravitational-wave emission from a binary made of the core of the donor star and the
companion star inspiraling within a shared common envelope.

Although common-envelope evolution is never \emph{driven} by the loss
of energy to gravitational radiation, the binary inside the shared
envelope does have a time-dependent mass quadrupole moment and can
thus emit gravitational radiation. If detectable, such gravitational
radiation offers invaluable insight on the process by passively
tracing the motion of the binary.

While the final dynamical phase of a common envelope might
emit detectable signals \citep{ginat:20}, its short duration makes
it an extremely rare target. Conversely, we focus on a longer
duration phase that might occur in some common-envelope events: the
self-regulating thermal-timescale phase. Detections of this
phase would constrain the stalling radius (from the gravitational-wave
frequency) and the duration of the stalled phase (from the number of
sources detected and/or the signal duration). Even non-detections
could put upper-limits on the duration or lower limits on the radius
at which common envelopes might stall (see Sec.~\ref{sec:non-detection}).

The gravitational-wave frequency range of interest is likely in the
mHz range. This is suggested by the existence of verification
white-dwarf binary for the LISA mission, which are thought to be the
outcome of successful common-envelope ejections.

Based on rapid population synthesis calculations, we estimate that
about one common-envelope event is initiated per decade in the Galaxy
(cf.~Fig.~\ref{fig:rates}), in agreement with previous observational
and theoretical determinations (Sec.~\ref{sec:rate} and
Sec.~\ref{sec:other_rates}). The largest uncertainty in converting
this to a number of Galactic sources is the duration of the common
envelope, which is predicted to lie between a dynamical ($\sim$ days)
and a thermal (decades to $10^5$\,years) timescale, and is likely to
depend on the masses and evolutionary stages of the binary
considered. Nevertheless, assuming that some common-envelope events go
through a self-regulated thermal-timescale phase, it is realistic to
expect at least one source within the LISA mission, and possibly more
than hundreds. 

While we remained agnostic on the details of the physics governing the
common-envelope dynamics, we can bracket the range of possibilities by
requiring the common envelope to proceed faster than if driven by pure
gravitational-wave emission (corresponding to the unrealistic
assumption of neglecting the gas drag), and slower than if the gas drag
was constant (corresponding to the unrealistic assumption of neglecting
the envelope reaction to the inspiraling binary). We found that a
system representative of the most common kind of Galactic common envelope
-- a $0.5\,M_\odot$ core with a $0.3\,M_\odot$ companion embedded in a
shared envelope -- might be (marginally) detectable during the stalled
phase within these limits (cf.~Fig.~\ref{fig:snrs}). More
specifically, if the separation at which the inspiral stalls is of
order of $0.1\,R_\odot$ and lasts $\gtrsim 5$\,years, a SNR $>$10 is
possible. The shared envelope is likely too low density to contribute
directly to the gravitational-wave signal.

We have further investigated the risk of misinterpreting a future
detection of a Galactic common-envelope binary if analyzed as a
compact binary in vacuum evolving according to general relativity
(cf.~Sec.~\ref{sec:bias}). Although the possibility of
misinterpretation exist (cf.~Fig.~\ref{fig:bias}), there are several
ways in which it could be avoided. To break the degeneracies and
recognize a signal as the product of a Galactic common envelope, one
could couple electromagnetic followup or using electromagnetic
observations to attempt directional searches in the gravitational-wave
data (Sec.~\ref{sec:EM_counterparts}); measure deviations from the
``braking index'' predicted by general relativity (for fast evolving
signals); or compare the rate of detections with the double
compact-object-merger rate from ground-based detectors.

Space-based gravitational-wave detectors might
improve our understanding the inner dynamics of
common-envelope evolution by using gravitational waves to probe in a
direct way a phenomenon that is \emph{not} gravitational wave driven.


\software{
  COSMIC \citep{breivik:20},
  \texttt{ipython/jupyter} \citep{ipython},
  \texttt{matplotlib} \citep{matplotlib},
  \texttt{NumPy} \citep{numpy}.
}

\acknowledgements{ Portions of this study were performed during the
  LISA Sprint at the Center for Computational Astrophysics of the
  Flatiron Institute, supported by the Simons Foundation. MR thanks
  S.~Justham and Y.~F.~Jiang for helpful discussions early on during
  this project, and K.~Breivik for guidance in using COSMIC and
  helpful feedback. We thank T.~Littenberg for useful discussion on
  LISA parameter estimation.}

\appendix

\section{Calculation of LISA Signal-to-Noise ratios}
\label{sec:snr}

We calculate expected LISA signal-to-noise ratios following the
discussion in \cite{robson:19}, our code is available at
\url{https://github.com/tcallister/LISA-and-CE-Evolution/}.
The gravitational-wave signal from a quasicircular binary can be
generically described via
    \begin{equation}
    \begin{aligned}
    h_+(t) &= A(t) \frac{1+\cos^2\iota}{2} \cos \Psi(t) \ \ ,\\
    h_\times(t) &= A(t) \cos\iota \sin \Psi(t).
    \end{aligned}
    \end{equation}
    Here, $A(t)$ is the gravitational wave's amplitude, $\iota$ is the
    inclination of the binary's orbital plane with respect to our line
    of sight, and $\Psi(t)$ is the gravitational wave's phase (the
    corresponding frequency of gravitational-wave emission is
    $f_{\rm GW} = \frac{1}{2\pi} \dot \Psi$).  In the frequency domain
    (denoted by a tilde), the resulting signal that is
    \textit{measured} by LISA is the linear combination
    \begin{equation}
    \tilde h(\fGW) = F_+(\fGW,\hat n)\,\tilde h_+(\fGW) + F_\times(\fGW,\hat n)\,\tilde h_\times(\fGW) \ \ ,
    \end{equation}
    assuming a signal incident on LISA from direction $\hat n$, and
    where $F_+(\fGW,\hat n)$ and $F_\times(\fGW,\hat n)$ are the LISA
    antenna response functions.  Note that, unlike ground-based
    detectors, LISA observes signals whose
    wavelengths are comparable to the size of the instrument itself,
    and therefore its antenna patterns are strongly
    frequency-dependent.  Also, here we explicitly use $\fGW$ to refer
    to gravitational-wave frequencies in order to minimize confusion
    with \textit{orbital} frequencies $f_\mathrm{orb} = \fGW/2$.

The resulting signal-to-noise ratio (SNR) with which this signal is observed in LISA is given by
\begin{equation}
\label{eq:snr-prelim}
{\rm SNR}^2 = 4 \int_0^\infty \frac{|h(\fGW)|^2}{S_n(\fGW)} d\fGW \ \ ,
\end{equation}
$S_n(\fGW)$ is the LISA strain sensitivity curve; we adopt the analytic expression in Eq.~1 of \cite{robson:19}.
As we do not know in advance the sky location or orientation of possible CE sources, Eq.~\eqref{eq:snr-prelim} can be averaged over all sky positions and binary inclination angles, yielding an \textit{expected} signal-to-noise ratio
\begin{equation}
\label{eq:snr-final}
\langle{\rm SNR}^2\rangle = \frac{16}{5} \int_0^\infty \frac{|\tilde A(\fGW)|^2}{S_n(\fGW)} d\fGW \ \ .
\end{equation}
We use Eq.~\eqref{eq:snr-final} to obtain the results shown with the
colors in Fig.~\ref{fig:snrs}.

To compute the Fourier domain amplitudes $\tilde A(\fGW)$, we first note that the time-domain amplitude of a quasi-circular binary with chirp mass $\mathcal{M}_c$, at distance $D$, and gravitational-wave frequency $\fGW$ is~\citep{maggiore_gravitational_2008}
\begin{equation}
A(t) = \frac{4}{D} \left(\frac{G\mathcal{M}_c}{c^2}\right)^{5/3} \left(\frac{\pi \fGW}{c}\right)^{2/3} \ \ .
\end{equation}
The corresponding Fourier-domain amplitude is~\citep{Buonanno_2009}
\begin{equation}
\tilde A(\fGW) = \frac{A\left( t(\fGW) \right)}{\sqrt{\dotfGW\left( t(\fGW) \right)}} \ \ ,
\end{equation}
where $\dotfGW$ is the time derivative of the gravitational-wave frequency and we now regard time as a function of $\fGW$.

\section{Calculation of gas-drag neglecting the envelope reaction}
\label{sec:gas_drag}

Any massive object moving linearly through gas with a constant density
interacts with its own gravitational wake and loses momentum via
dynamical friction. An object of mass $M$ and radius $R$ travelling
with velocity $v$ through gas of uniform density $\rho$ and
sound speed $c_{s}$ generates an enhanced density tail. This wake is
confined to the Mach cone at supersonic speeds
($\mathcal{M}=v/c_{s} \gg 1$) and at sub-sonic speeds the over-dense
wake lies within a sphere of radius $c_{s}t$ a distance $vt$ behind
the mass. If the wake is low amplitude, linear perturbation theory
yields a drag force of the form
$F_{\rm DF}=-C 4\pi (GM)^{2}\rho/v^{2}$ \citep{ostriker:99} where
$C$ is a term that approaches $C \rightarrow \mathcal{M}^{3}/3$ as
$\mathcal{M}$ \referee{$\ll$} $1$ and $C \rightarrow \ln(vt/R)$ as
$\mathcal{M}$ \referee{$\gg$} $1$. For a massive object on a circular orbit, the
path of the wake bends and is therefore asymmetric, leading to both
azimuthal and radial components of drag. The azimuthal component
dominates and is generally well approximated by $F_{\rm DF}$
\citep{kim:07}. Where the perturber is more massive, the wake is
larger amplitude and the non-linear regime modifies $F_{\rm DF}$ by
an additional factor $(\eta/2)^{-0.45}$ where $\eta$ is a non-linear
factor given by $\eta=A/(\mathcal{M}^{2}-1)$ and $A=GM/c_{s}^2R_{s}$
with $R_{s}$ a characteristic softening scale
\citep{kim:09}. Further modifications to the \citet{kim:07} picture
of drag can occur when the embedded mass is a binary and the binary
orbit or its orbit around its own center of mass become supersonic
\citep{sanchez:14}.

To calculate the solid orange line in Fig.~\ref{fig:snrs}, we assumed
$\rho=10^{-6}\,\mathrm{g\ cm^{-3}}$, roughly corresponding to the
average envelope density for a red supergiant, and a temperature
$T=\referee{9\times10^5}\,\mathrm{K}$, corresponding to the temperature somewhere
inside the envelope. To obtain a
(conservative) upper-limit on the gas drag force, we consider these parameters constant throughout
the inspiral, that is we neglect the reaction of the envelope to the
injection of energy and angular momentum by the binary spiraling-in.
With these parameters, we calculated $F_{\rm DF}$ and assumed it to be
the only source of energy loss and inspiral.

\section{Post-common-envelope separation with more massive companion}
\label{sec: CE-sep-massive-comp}
\begin{figure}[htbp]
    \centering
    \includegraphics[width=0.45\textwidth]{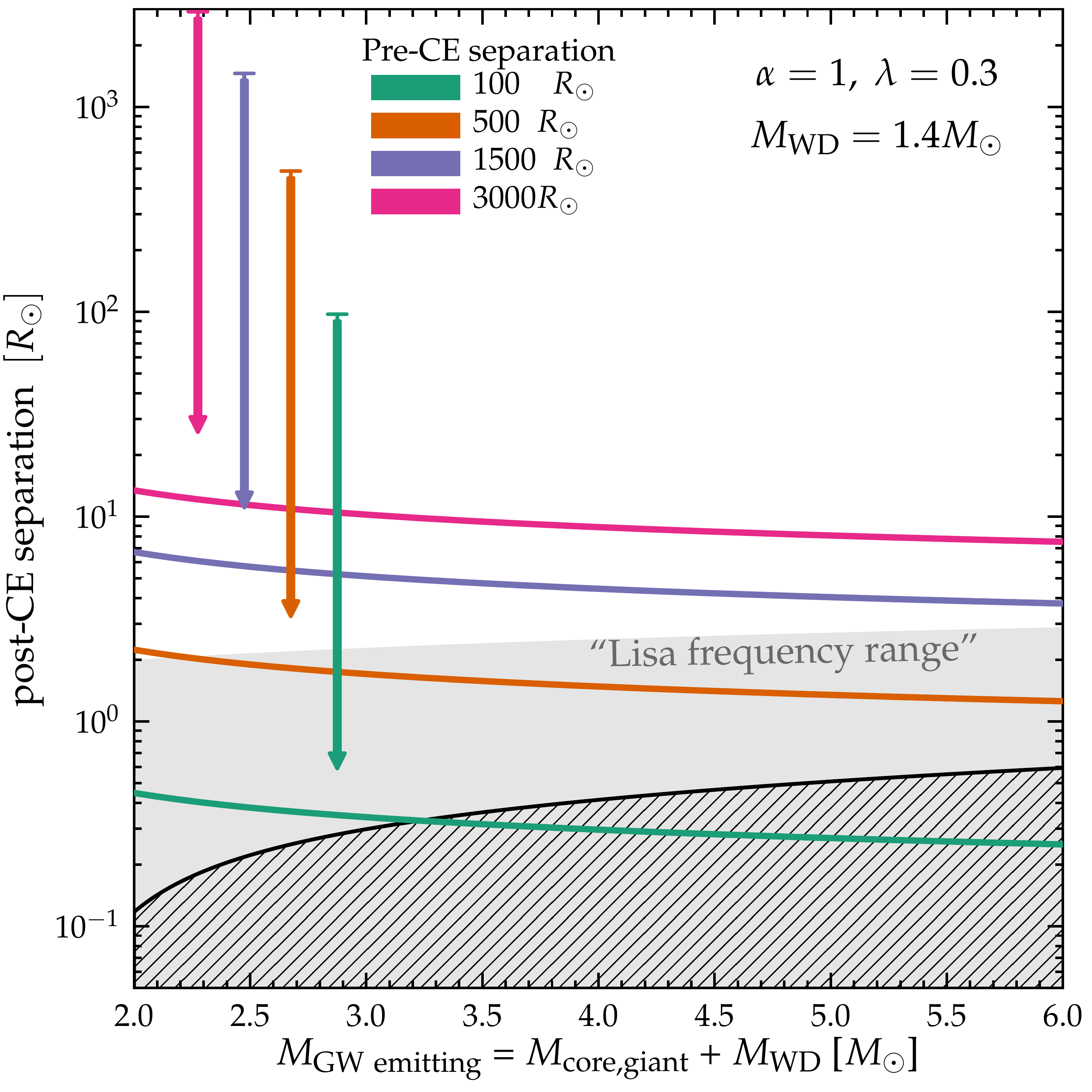}
    \caption{Same as Fig.~\ref{fig:CE_sep} but assuming a compact
      companion of $1.4\,M_\odot$ representing a Chandrasekhar mass WD
      or a neutron star.}
    \label{fig:CE_sep_1.4}
\end{figure}

\bibliography{./CE_LISA}

\end{document}